\documentclass[twocolumn, 10pt]{IEEEtran}
\usepackage{hyperref} 
\usepackage{graphicx}
\usepackage{amssymb}
\usepackage{amsmath}
\usepackage{mathtools}
\usepackage{cite}
\usepackage{stfloats}
\usepackage{subfigure}
\usepackage{epstopdf}
\usepackage{psfrag}
\usepackage[mathscr]{euscript}
\usepackage{acronym}  
\usepackage{booktabs} 
\usepackage[table]{xcolor}

\acrodef{CCDF}{complementary cumulative distribution function}
\acrodef{CF}{characteristic function}
\acrodef{PPP}{Poisson point process}
\acrodef{CSI}{channel state information}
\acrodef{OFDM}{orthogonal frequency division multiplexing}
\acrodef{OFDMA}{orthogonal frequency division multiple access}

\acrodef{RV}{random variable}
\acrodef{i.i.d.}{independent, identically distributed}
\acrodef{PMF}{probability mass function}
\acrodef{PDF}{probability distribution function}
\acrodef{CDF}{cumulative distribution function}
\acrodef{PGFL}{probability generating functional}
\acrodef{ch.f.}{characteristic function}
\acrodef{AWGN}{additive white Gaussian noise}
\acrodef{SNR}{signal-to-noise ratio}
\acrodef{LRT}{likelihood ratio test}
\acrodef{DRT}{distance ratio test}
\acrodef{GLRT}{generalized likelihood ratio test}
\acrodef{CRLB}{Cram\'{e}r-Rao lower bound}
\acrodef{CRB}{Cram\'{e}r-Rao bound}
\acrodef{ZZLB}{Ziv-Zakai lower bound}
\acrodef{ZZB}{Ziv-Zakai bound}
\acrodef{LoS}{line-of-sight}
\acrodef{ToF}{time-of-flight}
\acrodef{NLoS}{non-line-of-sight}
\acrodef{GDOP}{geometric dilution of precision}
\acrodef{GPS}{Global Positioning System}
\acrodef{FIM}{Fisher information matrix}
\acrodef{PEB}{position error bound}
\acrodef{SPEB}{squared position error bound}
\acrodef{TOA}{time-of-arrival}
\acrodef{TOF}{time-of-flight}
\acrodef{WSN}{wireless sensor network}
\acrodef{MAC}{medium access control}
\acrodef{RSS}{received signal strength}
\acrodef{WAF}{wall attenuation factor}
\acrodef{TDOA}{time difference-of-arrival}
\acrodef{RF}{radiofrequency}
\acrodef{RTT}{round-trip time}
\acrodef{AOA}{angle-of-arrival}
\acrodef{MF}{matched filter}
\acrodef{ED}{energy detector}
\acrodef{ML}{maximum likelihood}
\acrodef{MSE}{mean-square error}
\acrodef{RMSE}{root-mean-square error}
\acrodef{LEO}{localization error outage}
\acrodef{ppm}{part-per-million}
\acrodef{ACK}{acknowledge}
\acrodef{UWB}{Ultrawide bandwidth}
\acrodef{TNR}{threshold-to-noise ratio}
\acrodef{LS}{least squares}
\acrodef{IR-UWB}{impulse radio UWB}
\acrodef{FCC}{Federal Communications Commission}
\acrodef{TH}{time-hopping}
\acrodef{PPM}{pulse position modulation}
\acrodef{MUI}{multi-user interference}
\acrodef{PDP}{power delay profile}
\acrodef{BPZF}{band-pass zonal filter}
\acrodef{SIR}{signal-to-interference ratio}
\acrodef{RFID}{radio frequency identification}
\acrodef{WPAN}{wireless personal area network}
\acrodef{WWB}{Weiss-Weinstein bound}
\acrodef{DP}{direct path}
\acrodef{MF}{matched filter}
\acrodef{MMSE}{minimum-mean-square-error}
\acrodef{SBS}{serial backward search}
\acrodef{SBSMC}{serial backward search for multiple clusters}
\acrodef{NBI}{narrowband interference}
\acrodef{WBI}{wideband interference}
\acrodef{INR}{interference-to-noise ratio}
\acrodef{CR}{channel response}
\acrodef{CIR}{channel impulse response}
\acrodef{CR}{channel  response}
\acrodef{RADAR}{radar}
\acrodef{MUR}{Multistatic radar}
\acrodef{JBSF}{jump back and search forward}
\acrodef{HDSA}{high-definition situation-aware}
\acrodef{RRC}{root raised cosine}
\acrodef{ST}{simple thresholding}
\acrodef{BTB}{Bellini-Tartara bound}
\acrodef{P-Max}{$P$-Max}  
\acrodef{MIMO}{multiple-input multiple-output}
\acrodef{MAP}{maximum a posteriori}
\acrodef{FG}{factor graph}
\acrodef{OP}{outage probability}
\acrodef{WED}{wall extra delay}
\acrodef{RMS}{root mean square}
\acrodef{SPAWN}{sum-product algorithm over a wireless network}
\acrodef{MDD}{minimum distance distribution}
\acrodef{MAP}{maximum a posteriori probability}
\acrodef{PAR}{probabilistic association rule}
\acrodef{AP}{access point}
\acrodef{HD}{half-duplex}
\acrodef{FD}{full-duplex}
\acrodef{IC}{interference cancellation}
\acrodef{HDHN}{hybrid-duplex heterogeneous network}
\acrodef{TDD}{time-division duplexing}
\acrodef{FDD}{frequency-division duplexing}
\acrodef{SINR}{signal-to-interference-plus-noise ratio}
\acrodef{UAV}{unmanned aerial vehicle}
\acrodef{GCS}{ground control station}
\acrodef{LTE}{long term evolution}
\usepackage{color}
\usepackage{dsfont}
\usepackage{bbm}

\newcommand{\Pm}{P_\text{t}}
\newcommand{\PI}{P_\text{u}}
\newcommand{\Noise}{\sigma^2}

\newcommand{\om}{\ell_\text{r}}

\newcommand{\HL}{\overline{H_\text{L}}}
\newcommand{\HN}{\overline{H_\text{N}}}

\newcommand{\Ws}[2]{{W_{}^{}}} 

\newcommand{\Lw}{\lambda_{\text{e}}}

\newcommand{\zu}{z_\text{u}}

\newcommand{\dk}{d_\text{tu}}
\newcommand{\qt}{\boldsymbol{q}_\text{t}}
\newcommand{\qr}{\boldsymbol{q}_\text{r}}
\newcommand{\qe}{\boldsymbol{q}_{\text{e}_i}}
\newcommand{\qu}{\boldsymbol{q}_\text{u}}

\newcommand{\TSIR}[2]{{\tau_{}^{}}}

\newcommand{\PLoS}{p_{\text{L}}(d_c, \zu)}
\newcommand{\PNLoS}{p_{\text{N}}(d_c, \zu)}

\newcommand{\Pstp}{p_\text{s}(d_\text{r}, \zu)}
\newcommand{\Peve}{p_\text{e}(\zu)}
\newcommand{\Pse}{p_\text{se}(d_\text{r}, \zu)}

\DeclareMathAlphabet{\mathsf}{OML}{cmbr}{m}{it}

\newtheorem{lemma}{Lemma}
\newtheorem{corollary}{Corollary}

\newcommand{\bd}{\begin{description}}
\newcommand{\ed}{\end{description}}
\newcommand{\be}{\begin{enumerate}}
\newcommand{\ee}{\end{enumerate}}
\newcommand{\bi}{\begin{itemize}}
\newcommand{\ei}{\end{itemize}}
\newcommand{\bl}{\begin{list}}
\newcommand{\el}{\end{list}}
\newcommand{\bt}{\begin{tabbing}}
\newcommand{\et}{\end{tabbing}}

\acrodef{BS}{base station}
\acrodef{G2A}{ground-to-air}
\acrodef{A2G}{air-to-ground}
\acrodef{A2A}{Air-to-Air}
\acrodef{G2G}{ground-to-ground}
\acrodef{IoT}{Internet of things}

\begin{document}

\newcommand{\paperTitle}{Securing Communications with \\ Friendly Unmanned Aerial Vehicle Jammers}
 
 

\title{\paperTitle}

\author{
	\IEEEauthorblockN{
	Minsu Kim, Seongjun Kim, and 
	Jemin~Lee \vspace{-2mm}
	}\\[0.5em]
	%
\thanks{
	M.\ Kim, S.\ Kim, and J.\ Lee are with the Department of Information and
	Communication Engineering, Daegu Gyeongbuk Institute of Science and
	Technology, Daegu 42988, South Korea
	(e-mail: \texttt{ads5577@dgist.ac.kr}, \texttt{kseongjun@dgist.ac.kr}, \texttt{jmnlee@dgist.ac.kr}).
}
\thanks{
	The corresponding author is J. Lee. 
}
}

\maketitle 
%

%
%
\setcounter{page}{1}
\acresetall
\begin{abstract}
In this paper, 
we analyze the impact of a friendly \ac{UAV} jammer on \ac{UAV} communications in the presence of multiple eavesdroppers. We first present channel components determined by the \ac{LoS} probability between the friendly \ac{UAV} jammer and the ground device, and introduce different channel fadings for \ac{LoS} and \ac{NLoS} links.
We then derive the secrecy transmission probability satisfying both constraints of legitimate and wiretap channels. 
We also analyze the secrecy transmission probability in the presence of randomly distributed multiple friendly \ac{UAV} jammers.
Finally, we show the existence of the optimal \ac{UAV} jammer location, and the impact of the density of eavesdroppers, the transmission power of the \ac{UAV} jammer, and the density of \ac{UAV} jammers on the optimal location. 
%
\end{abstract}
%
\acresetall
%
%
\section{Introduction}\label{sec:Intro}
%
As an \ac{UAV} communication has several advantages such as the \ac{LoS} environment and their flexible mobility, many researchers have studied the use of \acp{UAV} as a communication device \cite{ZenZhaLim:16}.
Specifically, by using the relation between the \ac{LoS} probability and the distance-dependent path loss, the optimal positioning of \acp{UAV} has been mainly studied.
When the \ac{UAV} height increases, the link between the \ac{UAV} and the ground device forms the better link due to increasing \ac{LoS} signal, while the link distance increases.
Hence, several works optimized the \ac{UAV} height to improve the communication performance \cite{KimLee:19,KimLeeQue:20}.

%
In \ac{UAV} communications, the secrecy is also an important issue due to the broadcast nature of wireless channels. To overcome this, the physical layer security has recently emerged as an effective approach for communication secrecy \cite{Wyner:75,CheCooRen:17}. Different from terrestrial communications, in \ac{UAV} communications, the \ac{UAV} and the ground devices form \ac{LoS} links with higher probability, so malicious eavesdroppers as well as legitimate receivers can receive the signal from the transmitting \ac{UAV} with higher power.
%
%
%
Hence, the works in \cite{LiuLeeQuek:19,BaoYanHas:19} provided the optimal deployment and trajectory of \acp{UAV}, which improve the effect of the jamming signal to the eavesdroppers, while reducing the effect of the interfering signal to the receivers.
Specifically, the optimal \ac{UAV} height and the transmit power were presented to minimize the secrecy outage probability in \cite{LiuLeeQuek:19}.
The intercept probability and the ergodic secrecy rate were presented by considering the effect of the \ac{UAV} height and the transmit power in \cite{BaoYanHas:19}.
However, the works in \cite{LiuLeeQuek:19,BaoYanHas:19} did not consider the friendly jammer which can reduce the eavesdropping probability.

Recently, the friendly jammer has been considered in \cite{XuHeYan:15,QiLiChu:18,YaoXu:19,ZhoYeoChe:18} to improve the secrecy performance.
For the case of the friendly terrestrial jammer,
	the optimal secrecy guard zone radius was presented to maximize secrecy throughput in \cite{XuHeYan:15}.
Different from terrestrial communications, the \ac{UAV} and the ground devices form \ac{LoS} links with higher probability in \ac{UAV} communications.
	Hence, the friendly \ac{UAV} jammer can generally give stronger jamming signals to eavesdroppers than a terrestrial jammer
	by having \ac{LoS} links to eavesdroppers. Furthermore, the  friendly \ac{UAV} jammer can also be readily located to maximize the jamming efficiency as it has on-demand mobility. 
	Therefore, in recent works such as \cite{QiLiChu:18,YaoXu:19,ZhoYeoChe:18}, the friendly \ac{UAV} jammer has also been considered.
	Specifically,
the secrecy energy efficiency was presented to analyze the effect of the transmission power and the density ratio of transmitters to eavesdroppers in \cite{QiLiChu:18}.
The optimal \ac{UAV} height and the secrecy guard zone size were presented to maximize the secrecy transmission capacity in \cite{YaoXu:19}.
The optimal deployment and transmission power of the friendly \ac{UAV} jammer were provided to maximize the intercept probability security region in \cite{ZhoYeoChe:18}. 
%
However, the works in \cite{QiLiChu:18,YaoXu:19} focused on the effect of the density ratio of friendly \ac{UAV} jammers to eavesdroppers instead of the specific location of the friendly \ac{UAV} jammer.
The optimal location of a friendly \ac{UAV} jammer was presented in \cite{ZhoYeoChe:18},
but the channel fading for the \ac{A2G} channel was not considered.
In addition, the work in \cite{ZhoYeoChe:18} did not show the effect of the eavesdropper density on the optimal location of the friendly \ac{UAV} jammer.

%
%
%
In this paper, we present the effect of a friendly \ac{UAV} jammer on the secrecy transmission probability.
We consider channel fadings and components, affected by horizontal and vertical distances between the friendly \ac{UAV} jammer and the ground devices including eavesdroppers.
%
%
The main contributions of this paper can be summarized as follows:
\begin{itemize}
	\item we consider realistic channel model, determined by the \ac{LoS} probability between a friendly \ac{UAV} jammer and a ground device;
	\item we derive the secrecy transmission probability considering different channel fadings for \ac{LoS} and \ac{NLoS} links;
	\item we also analyze the secrecy transmission probability by considering multiple \ac{UAV} jammers, randomly distributed in the network; and
	\item we finally show the optimal location of the friendly \ac{UAV} jammer that maximizes the secrecy transmission probability according to the eavesdropper density and the transmission power of the friendly \ac{UAV} jammer.
\end{itemize}
%
%
%
%
%
%
%
%
%
%
\section{System Model}\label{sec:models}
In this section, we describe the \ac{UAV} network and the channel model, affected by horizontal and vertical distances between the friendly \ac{UAV} jammer and ground devices.
\subsection{Network Description}
\begin{figure}[t!]
	\begin{center}   
		{ 
			\includegraphics[width=0.8\columnwidth]{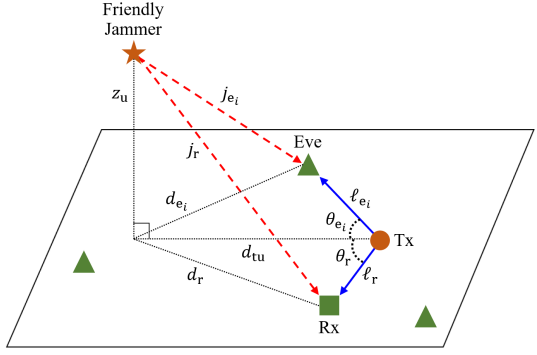}
			\vspace{-3mm}
		}
	\end{center}
	\caption{
		System model where multiple eavesdroppers are randomly distributed on the ground and a friendly UAV jammer is in the air. The blue lines represent transmitting signals from the transmitter and the red dotted lines represent interfering signals from the friendly \ac{UAV} jammer.
		\vspace{-2mm}
	}
	\label{fig:system}
\end{figure}

We consider the \ac{UAV} network with a transmitter (Tx), a legitimate receiver (Rx), a friendly \ac{UAV} jammer (Jammer), and multiple eavesdroppers (Eves) as shown in Fig. \ref{fig:system}. 
On the ground, the Tx and the Rx are located at $\qt=(0, 0, 0)$ and $\qr=(x_\text{r}, y_\text{r}, 0)$, respectively. Eves are randomly distributed by a \ac{PPP} $\Phi_\text{e} \triangleq \{\qe\}$ with density $\Lw$, where $\qe=(x_{\text{e}_i}, y_{\text{e}_i}, 0)$ is the location of an arbitrary Eve \cite{HaeGan:09}.
Each Eve decodes the received signal from the Tx independently, i.e., we consider non-colluding Eves.
The Jammer is placed in an adjustable location $\qu=(x_\text{u}, y_\text{u}, z_\text{u})$.
In this network, we assume the Tx and the Jammer do not know the locations of Eves. 
We also assume the legitimate channel (between Tx and Rx in the presence of Jammer) and the wiretap channel (between Tx and Eves in the presence of Jammer) are independent.\footnote{If the eavesdroppers are more than half-wavelength away from the legitimate users, the legitimate users experience independent channels to eavesdroppers \cite{Tse:05}.}
%
%
%
%
%

Based on $\qt$, $\qr$, $\qe$, and $\qu$, we define the \ac{SINR} of the legitimate channel $(c=\text{r})$ or the wiretap channel $(c=\text{e}_i)$ as
\begin{align}
	\gamma_c
	&=
	\frac{h_{\text{t}c} \ell_c^{-\alpha_{\text{t}c}} \Pm}  
	{h_{\text{u}c} j_c(d_c, \zu)^{-\alpha_{\text{u}c}} \PI + \Noise} \nonumber \\
	&=
	\frac{h_{\text{t}c} \rho_c}  
	{h_{\text{u}c} \tau_c(d_c, \zu) + \Noise}
	, \quad \quad c \in \{\text{r}, \text{e}_i\} \label{eq:SINR}
\end{align}
where 
$\ell_c = \sqrt{x_c^2 + y_c^2}$ is the link distance between the Tx and the Rx $(c=\text{r})$ or the Eve $(c=\text{e}_i)$, 
$j_c(d_c, \zu) = \sqrt{(x_\text{u} - x_c)^2 + (y_\text{u} - y_c)^2 + \zu^2}$ is the link distance between the Jammer and the Rx $(c=\text{r})$ or the Eve $(c=\text{e}_i)$,
$\Pm$ and $\PI$ are the transmission power of the Tx and the Jammer, respectively, and $\Noise$ is the noise power.
Here, $h_{\text{t}c}$ and $h_{\text{u}c}$ are channel fading gains, and $\alpha_{\text{t}c}$ and $\alpha_{\text{u}c}$ are path loss exponents, where the subscript $\text{t}c$ represents the transmission from the Tx to the Rx $(c=\text{r})$ or the Eve $(c=\text{e}_i)$, and the subscript $\text{u}c$ represents the transmission from the Jammer to the Rx $(c=\text{r})$ or the Eve $(c=\text{e}_i)$.\footnote{Note that the instantaneous \ac{CSI}, which is generally difficult to obtain especially for eavesdropping links, is not required in this work. Only the location of Tx and Rx as well as the densities of eavesdroppers and Jammers are needed to determine the location of Jammer.}
In \eqref{eq:SINR}, for convenience, we introduce $\rho_c=\ell_c^{-\alpha_{\text{t}c}} \Pm$ and $\tau_c(d_c, \zu)=j_c(d_c, \zu)^{-\alpha_{\text{u}c}} \PI$, where $d_c=\sqrt{(x_\text{u} - x_c)^2 + (y_\text{u} - y_c)^2}$ is horizontal distance between the Jammer and the Rx $(c=\text{r})$ or the Eve $(c=\text{e}_i)$, which can be expressed as
\begin{align}
d_c=\sqrt{\dk^2 + \ell_c^2 -2\dk\ell_c\cos{\theta_c}} \label{eq:horizontal_distance}
\end{align}
where $\dk$ is the horizontal distance between the Jammer and the Tx, and $\theta_c$ is the included angle between $\ell_c$ and $\dk$ as shown in Fig. \ref{fig:system}.
%
%
%
\subsection{Channel Model} \label{subsec:channel}
Since all the devices except for the Jammer are located on the ground, there can be two types of channels, which are
the \ac{G2G} channel and the \ac{A2G} channel.
The \ac{G2G} channel between ground devices is commonly modeled as the \ac{NLoS} environment with the Rayleigh fading due to a lot of obstacles.
On the other hand, the \ac{A2G} channel between the Jammer and the ground device (e.g., the interference link to Rx and the jamming link to Eve) can have the \ac{LoS} or \ac{NLoS} environment according to the existence of obstacles.
In this subsection, we introduce channel components and provide the model of the \ac{A2G} channel.
\subsubsection{Channel component} \label{subsec:component}
On \ac{UAV} communications, channel components such as the \ac{LoS} probability and the path loss exponent are affected by the horizontal distance $d_c$ and the vertical distance $\zu$ between a Jammer and a Rx $(c=\text{r})$ or a Eve $(c=\text{e}_i)$.
First, when the heights of ground devices are sufficiently small, the \emph{\ac{LoS} probability}, $\PLoS$, $c \in  \{\text{r}, \text{e}_i\}$, is given by \cite{YanZhoZha:18}\footnote{The \ac{LoS} probability is also defined in \cite{HouKanLar:14}, but it is determined by the ratio of the vertical and horizontal link distances, not by the absolute distances. On the other hand, the one in \cite{YanZhoZha:18} is affected by the absolute positions of the Tx and Rx, and it can be applicable for more general cases.}
\begin{align} \label{eq:LoS}
\PLoS
=
\left\{ 1 - \frac{\sqrt{2\pi} \zeta} {\zu} 
	\Bigg| 
			Q\left(\frac{\zu} {\zeta}\right) - 0.5
	\Bigg| \right\}^{d_c \sqrt{\nu \mu} }
\end{align}
where $Q(x)\hspace{-0.5mm}=\hspace{-0.5mm}\int_{x}^{\infty} \frac{1} {\sqrt{2\pi}} \exp \left(- \frac{t^2} {2}\right) \,dt$ is the Q-function, and $\zeta$, $\nu$, and $\mu$ are environment parameters, determined by the building density and height.
We then define the \ac{NLoS} probability as $\PNLoS=1-\PLoS$.
Note that the \ac{LoS} probability can be applicable in various environments (e.g., urban, suburban, and dense urban) by adjusting the environment parameters.
In addition, the \emph{path loss exponent} $\alpha_{\text{u}c}$ is determined by $\alpha_\text{L}$
when the \ac{A2G} channel is in the \ac{LoS} environment.
Otherwise, $\alpha_{\text{u}c}=\alpha_\text{N}$.
%
%
%
%
%
\subsubsection{Air-to-Ground (A2G) channel}
In the \ac{A2G} channel, as the received signal power at the ground device is affected by the combination of the \ac{LoS} and \ac{NLoS} signals \cite{Shi:06}, we consider that the channel fading is the Nakagami-$m$ fading with mean $\HL=1$ for the \ac{LoS} environment and the Rayleigh fading with mean $\HN=1$ for the \ac{NLoS} environment.
Therefore, the distribution of channel fading gains, $h_{\text{u}c}$, $c \in  \{\text{r}, \text{e}_i\}$, can be expressed as
\begin{align}	
f_{h_{\text{u}c}}^{(\text{L})}(h)
& = \frac{m_\text{L}^{m_\text{L}} h^{m_\text{L}-1}} {\Gamma\left(m_\text{L}\right)} \exp\left(-m_\text{L} h\right) \quad  \text{for \ac{LoS}}  \nonumber \\
f_{h_{\text{u}c}}^{(\text{N})}(h)
& = 
\exp\left(-h\right) \quad \quad \quad \quad \quad \,\,\, \quad \quad \text{for \ac{NLoS}} \label{eq:fading}
\end{align}
where $m_\text{L}$ is the Nakagami-$m$ fading parameter and $\Gamma(z)=\int_{0}^{\infty} x^{z-1} e^{-x} dx$ is the gamma function.
%
%
%
\section{Secrecy Transmission Probability Analysis}\label{sec:analytical}
In this section, for given $d_\text{r}$ and $\zu$,
we analyze the secrecy transmission probability $\Pse$, which is the probability that a Tx reliably transmits signals to a Rx, while all the Eves fail to eavesdrop, and is defined as \cite{XuHeYan:15}
\begin{align}
\Pse 
&= 
\mathbb{P}\left[\gamma_\text{r} > \gamma_{\text{t}}, 
\max\limits_{\text{e}_i \in \Phi_\text{e}}  \gamma_{\text{e}_i}  <  \gamma_{\text{t}}' \right]  \label{eq:p_se}
\end{align}
where $\gamma_{\text{t}}$ and $\gamma_{\text{t}}'$ are target \acp{SINR} of the legitimate channel and the wiretap channel, respectively.
\begin{lemma} \label{lemma:Pse}
The secrecy transmission probability can be presented as
\begin{align} \label{eq:secrecy}
\Pse = \Pstp (1 - \Peve) 
\end{align}
where $\Pstp$ and $\Peve$ are given by
\begin{align}
&\Pstp \hspace{-0.5mm} = \hspace{-0.5mm}
\frac{m_\text{L}^{m_\text{L}}} {\left(\frac{\gamma_\text{t} \tau_\text{r}(d_\text{r}, \zu)} {\rho_\text{r}} + m_\text{L}\right)^{m_\text{L}}} 
\exp  \left( - \frac{\gamma_\text{t} \Noise} {\rho_\text{r}} \right)  p_\text{L}(d_\text{r},\zu) \nonumber \\
& \quad\quad\quad\quad\,\,\, +
\frac{\rho_\text{r}}  
{\rho_\text{r} + \gamma_\text{t} \tau_\text{r}(d_\text{r},\zu)}
\exp \left( -\frac{\gamma_\text{t} \Noise}  {\rho_\text{r}}\right) p_\text{N}(d_\text{r},\zu),  \label{eq:P_STP}\\
&
p_\text{e}\hspace{-0.2mm}(\hspace{-0.2mm}\zu\hspace{-0.2mm}) 
\hspace{-0.5mm} = \hspace{-0.5mm}
1 \hspace{-0.5mm} - \hspace{-0.2mm} \exp \hspace{-0.5mm}
	\left\{\hspace{-0.5mm} - \Lw \hspace{-1mm} \int_{0}^{2 \pi} \hspace{-2.3mm} \int_{0}^{\infty} \hspace{-0.5mm} p_{\text{s,e}_i}\hspace{-0.3mm}(\hspace{-0.2mm}\ell_{\text{e}_i}\hspace{-0.2mm}, \hspace{-0.2mm} \theta_{\text{e}_i}\hspace{-0.2mm}, \hspace{-0.2mm}\zu\hspace{-0.1mm}) \ell_{\text{e}_i} d\ell_{\text{e}_i} d\theta_{\text{e}_i} \hspace{-0.3mm} \right\} \hspace{-0.7mm} . \hspace{-1mm} \label{eq:P_EVE}
\end{align}
In \eqref{eq:P_EVE}, $p_{\text{s,e}_i}(\ell_{\text{e}_i}, \theta_{\text{e}_i}, \zu) = \mathbb{P}\left[\gamma_{\text{e}_i} > \gamma_{\text{t}}'\right]$, which is presented from \eqref{eq:P_STP} by substituting from $d_\text{r}$, $\rho_\text{r}$, and $\gamma_\text{t}$ to $d_{\text{e}_i}$, $\rho_{\text{e}_i}$, and $\gamma_\text{t}'$, respectively.\footnote{For given $\zu$, even though the eavesdropping probability $\Peve$ cannot be obtained in a closed-form, we can calculate it efficiently using the numerical integral method.}
\end{lemma}
\begin{IEEEproof}
	For given $\qt$, $\qu$, $\qe$, and $\qr$, we can obtain the secrecy transmission probability $\Pse$ as
	\begin{align}
	\Pse 
	&= 
	\mathbb{P}\left[\gamma_\text{r} > \gamma_{\text{t}}, 
	\max\limits_{\text{e}_i \in \Phi_\text{e}}  \gamma_{\text{e}_i}  <  \gamma_{\text{t}}' \right]  \nonumber \\
	&\overset{\underset{\mathrm{(a)}}{}}{=} \Pstp (1 - \Peve) \label{eq:lemma}
	\end{align}
	where $\Pstp = \mathbb{P}\left[\gamma_\text{r} > \gamma_{\text{t}}\right]$ is the successful transmission probability, $\Peve = \mathbb{P}\left[\max\limits_{\text{e}_i \in \Phi_\text{e}}  \gamma_{\text{e}_i}  >  \gamma_{\text{t}}'\right]$ is the eavesdropping probability, and (a) is obtained due to the independence between the legitimate channel and the wiretap channel.
	In \eqref{eq:lemma}, $\Pstp$ can be obtained as \cite{KimLee:19}
	\begin{align}
	&\Pstp 	= 
	\int_{0}^{\infty}
	\int_{\frac{\gamma_\text{t} \left(\tau_\text{r}(d_\text{r},\zu) g  +  \Noise \right) } {\rho_\text{r}}}^{\infty}  
	%
	%
	f_{h_\text{tr}}(h) dh  f_{h_\text{ur}}(g)  dg \nonumber \\
	&= 
	p_\text{s}^{(\text{L})}(d_\text{r}, \zu) p_\text{L}(d_\text{r}, \zu) 	
	+ 
	p_\text{s}^{(\text{N})}(d_\text{r}, \zu) 
	p_\text{N}(d_\text{r}, \zu)  \label{eq:STP}
	\end{align}
	where $p_\text{s}^{(e_\text{I})}(d_\text{r}, \zu)$ is the successful transmission probability in the environment of the interference link $e_\text{I}$, given by
	\begin{align}
	&p_\text{s}^{(\text{L})}(d_\text{r}, \zu)
	= 
	\int_{0}^{\infty} 
	\int_{\frac{\gamma_\text{t} \left(\tau_\text{r}(d_\text{r}, \zu) g + \Noise \right) } {\rho_\text{r}}}^{\infty}  
	%
	%
	f_{h_\text{tr}}(h) \, dh  f_{h_\text{ur}}^{(\text{L})}(g)  \, dg \nonumber \\
	&\overset{\underset{\mathrm{(a)}}{}}{=} \hspace{-1.2mm}
	\int_{0}^{\infty}  \hspace{-1.5mm}
	\exp \hspace{-0.5mm}
		\left\{ \hspace{-0.5mm}
		-\frac{\gamma_\text{t} \hspace{-0.7mm} \left(\hspace{-0.2mm}\tau_\text{r}(d_\text{r}\hspace{-0.2mm},\hspace{-0.3mm} \zu\hspace{-0.2mm}) g \hspace{-0.7mm} + \hspace{-0.7mm} \Noise \right)} {\rho_\text{r}}  
		\hspace{-0.5mm} - \hspace{-0.5mm} m_\text{L} g \hspace{-0.5mm}
		\right\} \hspace{-0.5mm}
	\frac{m_\text{L}^{m_\text{L}} \hspace{-0.2mm} g^{m_\text{L}\hspace{-0.2mm}-\hspace{-0.2mm}1}} {\Gamma(m_\text{L})} dg , \hspace{-1.5mm} \label{eq:STP_L}\\
	&p_\text{s}^{(\text{N})}(d_\text{r}, \zu)
	= 
	\int_{0}^{\infty} 
	\int_{\frac{\gamma_\text{t} \left(\tau_\text{r}(d_\text{r}, \zu) g + \Noise \right) } {\rho_\text{r}}}^{\infty}  
	%
	%
	f_{h_\text{tr}}(h) \, dh  f_{h_\text{ur}}^{(\text{N})}(g)  \, dg \nonumber \\
	&\overset{\underset{\mathrm{(a)}}{}}{=}
	\int_{0}^{\infty} 
	\exp 
	\left\{ -\frac{\gamma_\text{t} \left(\tau_\text{r}(d_\text{r}, \zu) g + \Noise \right)} {\rho_\text{r}} - g  \right\} dg . \label{eq:STP_N}
	\end{align} 
	%
	%
	Here, (a) is from the \ac{CDF} of the exponential distribution.
	In \eqref{eq:STP_L}, by \cite[eq. (3.326)]{GraRyz:B07}, 
	the integral term can be expressed as
	\begin{align}
		\int_{0}^{\infty} x^m \exp\left(-\beta x^n\right) dx
		= \frac{\Gamma(\gamma)}{n \beta^\gamma} \label{eq:bessel}
	\end{align}
	where $m=m_\text{L}-1$, $n=1$, $\beta = \frac{\gamma_\text{t} \tau_\text{r}(d_\text{r},\zu)} {\rho_\text{r}} + m_\text{L}$, and $\gamma=\frac{m+1}{n}$.
	By using \eqref{eq:bessel} in \eqref{eq:STP_L} and the definite integral in \eqref{eq:STP_N}, $\Pstp$ is presented as \eqref{eq:P_STP}.
	
	In the wiretap channel, $\Peve$ can be derived as
	\begin{align} \label{eq:outage}
	\Peve  \hspace{-0.3mm} = \hspace{-0.3mm}
	\mathbb{P} \hspace{-0.3mm} \left[\max\limits_{\text{e}_i \in \Phi_\text{e}}  \gamma_{\text{e}_i} \hspace{-0.5mm} > \hspace{-0.3mm}  \gamma_{\text{t}}'\right] 
	\hspace{-0.6mm} = \hspace{-0.6mm}	1 \hspace{-0.3mm} - \hspace{-0.3mm} \mathbb{E}_{\Phi_\text{e}} \hspace{-0.5mm}
	\left[\prod_{\text{e}_i \in \Phi_\text{e}} \hspace{-0.3mm}
	\mathbb{P} \hspace{-0.2mm} \left[\gamma_{\text{e}_i} \hspace{-0.5mm} < \hspace{-0.3mm} \gamma_{\text{t}}'\right]\right] \hspace{-0.7mm} .  \hspace{-1mm}
	\end{align}
	By using the \ac{PGFL} in \eqref{eq:outage}, $\Peve$ is presented as \eqref{eq:P_EVE}.
\end{IEEEproof}

From Lemma \ref{lemma:Pse}, we can know that $\Pstp$ and $\Peve$ decrease with $m_\text{L}$.
Using this result, the impact of $m_\text{L}$ on $p_\text{se}(d_\text{r},\zu)$ is shown and discussed in the numerical results.
%
%
%
%
%
\begin{corollary} \label{cor:theta_r}
	For given $\zu$, $\ell_{r}$, and $\dk$, the optimal value of $\theta_\text{r}$ that maximizes $p_\text{se}(\dk,\zu,\theta_\text{r})$ is $\pi$.
\end{corollary}
%
%
%
%
\begin{IEEEproof}
For convenience, we introduce $F(\dk,\zu)=\int_{0}^{2\pi} \hspace{-1mm} \int_{0}^{\infty}  p_{\text{s,e}_i}(\ell_{\text{e}_i}, \theta_{\text{e}_i}, \zu) \ell_{\text{e}_i} d\ell_{\text{e}_i} d\theta_{\text{e}_i}$ and represent $\Pse$ as functions of $\dk$, $\zu$, and $\theta_\text{r}$ as
\begin{align}
p_\text{se}(\dk,\zu,\theta_\text{r}) = p_\text{s}(\dk,\zu,\theta_\text{r}) \exp\left\{-\Lw F(\dk,\zu)\right\}.
\label{eq:secrecy_re}
\end{align} 
In \eqref{eq:secrecy_re}, for given $\zu$, $\ell_{r}$, and $\dk$, we obtain the first derivative of $p_\text{se}(\dk,\zu,\theta_\text{r})$ with respect to $\theta_\text{r}$ as
\begin{align}
&\frac{\partial p_\text{se} (\dk,\zu,\theta_\text{r})} {\partial \theta_\text{r}} = 
\exp\left\{-\Lw F(\dk,\zu)\right\} 
\left\{ \frac{\partial p_\text{L} (\dk,\zu,\theta_\text{r})} {\partial \theta_\text{r}}  \right. \nonumber \\
& \left. \times
\left(p_\text{s}^{(\text{L})} (\dk,\zu,\theta_\text{r}) - p_\text{s}^{(\text{N})} (\dk,\zu,\theta_\text{r})\right) +
\frac{\partial p_\text{s}^{(\text{L})} (\dk,\zu,\theta_\text{r})} {\partial \theta_\text{r}} \right. \nonumber \\
& \left. \times
p_\text{L} (\dk,\zu,\theta_\text{r}) 
+ \frac{\partial p_\text{s}^{(\text{N})} (\dk,\zu,\theta_\text{r})} {\partial \theta_\text{r}} 
p_\text{N} (\dk,\zu,\theta_\text{r})
\right\}.
\label{eq:derivative_theta}
\end{align}
In \eqref{eq:derivative_theta}, for $\zu>0$, $p_\text{s}^{(\text{L})}(\dk,\zu,\theta_\text{r}) \hspace{-0.1mm} - \hspace{-0.1mm} p_\text{s}^{(\text{N})} (\dk,\zu,\theta_\text{r}) \hspace{-0.2mm} < 0$, $p_\text{L}(\dk,\zu,\theta_\text{r}) \hspace{-0.2mm} > 0$, and $p_\text{N}(\dk,\zu,\theta_\text{r}) > 0$.
In addition, from \eqref{eq:LoS}, we obtain $\frac{\partial p_\text{L} (\dk,\zu,\theta_\text{r})} {\partial \theta_\text{r}} = -C_1 \sin(\theta_\text{r})$, $\frac{\partial p_\text{s}^{(\text{L})} (\dk,\zu,\theta_\text{r})} {\partial \theta_\text{r}} = C_2 \sin(\theta_\text{r})$, and $\frac{\partial p_\text{s}^{(\text{N})} (\dk,\zu,\theta_\text{r})} {\partial \theta_\text{r}} = C_3 \sin(\theta_\text{r})$ for positive $C_1$, $C_2$, and $C_3$.
Hence, the stationary values of $\theta_\text{r}$ are obtained when $\sin(\theta_\text{r})=0$.
Furthermore, we readily know that $p_\text{s}(\dk, \zu, \pi)$ is greater than $p_\text{s}(\dk, \zu, 0)$ because $p_\text{L} (\dk,\zu,\pi)$ is smaller than $p_\text{L} (\dk,\zu,0)$ and $\tau_\text{r}(\dk,\zu,\pi)$ is smaller than $\tau_\text{r}(\dk,\zu,0)$.
Therefore, the optimal value of $\theta_\text{r}$ that maximizes $p_\text{se}(\dk,\zu,\theta_\text{r})$ is $\pi$.
\end{IEEEproof}
%
%

From Corollary~\ref{cor:theta_r}, we can see that the Jammer needs to be located along the line from the Rx to the Tx.
Hence, in Section \ref{sec:numerical}, we analyze $p_\text{se}(\dk, \zu, \pi)$ instead of $p_\text{se}(\dk, \zu, \theta_{\text{r}})$.

We now present the asymptotic secrecy transmission probability when the Jammer is located near to the Tx.

\begin{corollary} \label{pro:asymptotic}
	As the Jammer approaches to the Tx, the asymptotic secrecy transmission probability can be given by
	\begin{align}
	\Pse \hspace{-0.3mm} \approx \hspace{-0.3mm}
	\Pstp \hspace{-0.3mm}
	\exp \hspace{-0.5mm} \left\{ \hspace{-0.5mm}
	\frac{-2 \Lw \pi P_\text{t} \Gamma\left(\frac{2}{\alpha_\text{N}}\right)} {\left(\gamma_\text{t}' P_\text{u} \hspace{-0.5mm} + \hspace{-0.5mm} P_\text{t}\right) \alpha_\text{N} \hspace{-0.5mm} \left(\frac{\gamma_\text{t}' \Noise} {P_\text{t}}\right)^\frac{2}{\alpha_\text{N}}} \hspace{-0.5mm}
	\right\} \hspace{-0.5mm} \label{eq:asymptotic} 
	\end{align}
	where $\Pstp$ is given in \eqref{eq:P_STP}.
\end{corollary}
\begin{IEEEproof}
	In \eqref{eq:P_EVE}, as $d_\text{tu} \rightarrow 0$ (i.e., when the Jammer approaches to the Tx with the height $\zu$), the eavesdropping probability $\Peve$ can be given by
	\begin{align}
	p_\text{e}(\zu) &\approx
	1 - \exp \left[ 
	- 2 \pi \Lw \int_{0}^{\infty} \left\{
	\exp \left( - \frac{\gamma_\text{t}' \Noise} {\rho_{\text{e}_i}}  \right)
	  \right. \right. \nonumber \\
	& \left. \left. \quad \times 
		\frac{m_\text{L}^{m_\text{L}} p_\text{L}(\ell_{\text{e}_i},\zu)} {\left(\frac{\gamma_\text{t}' P_\text{u} \left(\ell_{\text{e}_i}^2 + \zu^2\right)^{-\frac{\alpha_{\text{L}}}{2}}} {\rho_{\text{e}_i}} + m_\text{L}\right)^{m_\text{L}}} 
	+ \exp \left( - \frac{\gamma_\text{t}' \Noise} {\rho_{\text{e}_i}}  \right)
		\right. \right.  \nonumber \\
	& \left. \left. \quad \times
		\frac{\rho_{\text{e}_i} p_\text{N}(\ell_{\text{e}_i},\zu)}  
	{\rho_{\text{e}_i} + \gamma_\text{t}' P_\text{u} 
		\left(\ell_{\text{e}_i}^2 + \zu^2 \right)^{-\frac{\alpha_{\text{N}}}{2}}}
	\right\} \ell_{\text{e}_i} d\ell_{\text{e}_i} \right] .  \label{eq:EVE_asy}
	\end{align}
	In \eqref{eq:EVE_asy}, when $\zu$ is small, $p_\text{L}(\ell_{\text{e}_i},\zu)$ approaches to zero and $p_\text{e}(\zu)$ can be represented as
	\begin{align}
	p_\text{e} \hspace{-0.8mm} \approx \hspace{-0.8mm}
	1 \hspace{-0.7mm} - \hspace{-0.5mm} \exp \hspace{-0.8mm} 
	\left\{ \hspace{-0.7mm}
	- 2 \pi \hspace{-0.2mm} \Lw \hspace{-1.2mm} \int_{0}^{\infty} \hspace{-1.5mm}
	\frac{P_\text{t}}  
	{P_\text{t} \hspace{-0.8mm} + \hspace{-0.8mm} \gamma_\text{t}' P_\text{u}}
	\hspace{-0.5mm} \exp \hspace{-1mm} \left( \hspace{-1mm} -\frac{\gamma_\text{t}' \Noise}  {P_\text{t} \ell_{\text{e}_i}^{-\alpha_\text{N}}} \hspace{-0.7mm}\right) \hspace{-0.8mm} \ell_{\text{e}_i}  d\ell_{\text{e}_i} \hspace{-0.7mm} \right\} \hspace{-0.8mm} . \hspace{-1.5mm} \label{eq:P_EVE_A}
	\end{align}
	Using the following result \cite[eq. (3.326)]{GraRyz:B07}
	\begin{align}
	\int_{0}^{\infty} x^m \exp\left(-\beta x^n\right) dx
	= \frac{\Gamma(\gamma)}{n \beta^\gamma} \label{eq:integral}
	\end{align}
	with $m=1$, $n=\alpha_\text{N}$, $\beta=\frac{\gamma_\text{t}' \Noise}{P_t}$, and $\gamma=\frac{m+1}{n}$,
	$p_\text{e}$ in \eqref{eq:P_EVE_A} can be presented in a closed-form. 
	Finally, we can obtain the asymptotic expression of $\Pse$ as \eqref{eq:asymptotic}.
\end{IEEEproof}

From Corollary~\ref{pro:asymptotic}, we can readily see the effect of the network parameters (e.g., main link distance and transmission power of Tx) on the secrecy transmission probability.

\section{Secrecy Transmission Probability Analysis With Multiple UAV Jammers}

In this section, we now consider multiple \ac{UAV} jammers, which are randomly distributed by a \ac{PPP} $\Phi_\text{u}$ with density $\lambda_\text{u}$ at the height $\zu$.
The channel components and fading gains between the typical Jammer and the ground device are the same as the single Jammer case.
From the secrecy transmission probability in \eqref{eq:secrecy}, we obtain the secrecy transmission probability for multiple \ac{UAV} jammers in the following corollary.

%
%
\begin{corollary} \label{cor:multiple}
	In the presence of multiple \ac{UAV} jammers, the secrecy transmission probability $\tilde{p}_\text{se}(\zu)$ is given by
	\begin{align}
	&\tilde{p}_\text{se}\hspace{-0.2mm}(\hspace{-0.3mm}\zu\hspace{-0.2mm}) \hspace{-0.8mm} = \hspace{-0.6mm}
	\exp \hspace{-0.5mm} \left\{ \hspace{-0.5mm}
		- 2 \pi \lambda_\text{u} \hspace{-1.2mm} \int_{0}^{\infty} \hspace{-1.3mm}
		 \left(\hspace{-0.5mm} 1 \hspace{-0.6mm} - \hspace{-1.4mm} 
		 \sum_{e_\text{I} \in \{\text{L}, \text{N}\}} \hspace{-1.2mm}
		\hat{p}_\text{s}^{(e_\text{I})}\hspace{-0.3mm}(\hspace{-0.3mm}r,\hspace{-0.3mm}\zu\hspace{-0.2mm})p_{e_\text{I}}(\hspace{-0.3mm}r,\hspace{-0.3mm}\zu\hspace{-0.2mm}) \hspace{-0.9mm} \right) \hspace{-0.8mm} r dr \hspace{-0.5mm} \right. \nonumber \\
	& \left.
	- \frac{\gamma_\text{t} \Noise} {\rho_\text{r}} \right\}
	\exp\left[- 2 \pi \Lw \int_{0}^{\infty} 
	\exp\left\{
	- \frac{\gamma_\text{t}' \Noise} {\rho_{\text{e}_i}} - 2 \pi \lambda_\text{u} \right. \right.  \nonumber \\
	& \left. \left.  \times
	\int_{0}^{\infty} \hspace{-1mm} \left( 1 - \hspace{-1mm}
	\sum_{e_\text{J} \in \{\text{L}, \text{N}\}} \hspace{-1mm}
	\hat{p}_{\text{s,e}_i}^{(e_\text{J})}(v, \zu) p_{e_\text{J}}(v,\zu) \hspace{-0.5mm} \right) \hspace{-0.5mm} v dv \hspace{-0.5mm} \right\} \hspace{-0.5mm} \ell_{\text{e}_i} d\ell_{\text{e}_i} \hspace{-0.5mm} \right] \hspace{-2mm} \label{eq:p_se_multiple}
	\end{align}
	where $\hat{p}_\text{s}^{(e_\text{I})}(r,\zu)$ is the successful transmission probability of the interference limited environment (i.e., $\Noise=0$ in \eqref{eq:STP_L} and \eqref{eq:STP_N}) and $r$ is the horizontal distance between the Jammer and the Rx.
	In \eqref{eq:p_se_multiple}, $\hat{p}_{\text{s,e}_i}^{(e_\text{J})}(v, \zu)$ is obtained from $\hat{p}_\text{s}^{(e_\text{I})}(r,\zu)$ by replacing $d_{\text{u}_i\text{r}}$, $\rho_\text{r}$, and $\gamma_\text{t}$ with $d_{\text{u}_i{\text{e}_i}}$, $\rho_{\text{e}_i}$, and $\gamma_\text{t}'$, respectively.
	Here, $e_\text{J}$ is the environment of the jamming link and $v$ is the horizontal distance between the Jammer and the Eve.
\end{corollary}
%
%
%
%
\begin{IEEEproof}
	For given $\zu$, the secrecy transmission probability for multiple \ac{UAV} jammers can be presented as
	\begin{align}
	&\tilde{p}_\text{se}(\zu) = 
	\mathbb{E}_{\Phi_\text{u}} \left[\mathbb{P}\left[
	\frac{h_{\text{tr}} \rho_\text{r}}  
	{\sum_{\text{u}_i \in \Phi_\text{u}} h_{\text{u}_i\text{r}} \tau_\text{r}(d_{\text{u}_i\text{r}}, \zu) + \Noise} > \gamma_{\text{t}} \right] \right] \nonumber \\
	& \quad \times
	\mathbb{E}_{\Phi_\text{u}} \left[\mathbb{P}\left[
	\max\limits_{\text{e}_i \in \Phi_\text{e}}  \frac{h_{\text{te}_i} \rho_{\text{e}_i}}  
	{\sum_{\text{u}_i \in \Phi_\text{u}} h_{\text{u}_i\text{e}_i} \tau_{\text{e}_i}(d_{\text{u}_i{\text{e}_i}}, \zu) + \Noise}  <  \gamma_{\text{t}}' \right] \right] \nonumber \\
	& \overset{\underset{\mathrm{(a)}}{}}{=}
	\mathbb{E}_{\Phi_\text{u}} \left[ \prod_{\text{u}_i \in \Phi_\text{u}} 
	\mathbb{E}_{h_{\text{u}_i\text{r}}} \left[\exp \left(- \frac{\gamma_\text{t} h_{\text{u}_i\text{r}} \tau_\text{r}(d_{\text{u}_i\text{r}}, \zu)} {\rho_\text{r}} \right) \right] \right] \nonumber \\
	& \quad \times
	\exp \left( - \frac{\gamma_\text{t} \Noise} {\rho_\text{r}} \right)
	\mathbb{E}_{\Phi_\text{e}} 
	\left[\prod_{\text{e}_i \in \Phi_\text{e}}  \left\{ 1 -
	 \exp \left( - \frac{\gamma_\text{t}' \Noise} {\rho_{\text{e}_i}} \right) \right. \right. \nonumber \\
	 & \quad \left. \left. \times
	\mathbb{E}_{\Phi_\text{u}} 
	\left[	\prod_{\text{u}_i  \in \Phi_\text{u}} 
	\mathbb{E}_{h_{\text{u}_i{\text{e}_i}}} \left[\exp  \left( - \frac{\gamma_\text{t}' h_{\text{u}_i{\text{e}_i}} \tau_{\text{e}_i}(d_{\text{u}_i{\text{e}_i}}, \zu)} {\rho_{\text{e}_i}}  \right) \right] \right] \right\} \right]  \nonumber \\
	& \overset{\underset{\mathrm{(b)}}{}}{=}
	\mathbb{E}_{\Phi_\text{u}} \hspace{-1.2mm} \left[ \prod_{\text{u}_i \in \Phi_\text{u}} 
	\sum_{e_\text{I} \in \{\text{L}, \text{N}\}}  \hspace{-1.5mm}
	\hat{p}_\text{s}^{(\hspace{-0.2mm}e_\text{I}\hspace{-0.2mm})} \hspace{-0.3mm}(\hspace{-0.2mm}d_{\text{u}_i\text{r}}, \hspace{-0.3mm} \zu\hspace{-0.2mm}) p_{e_\text{I}}\hspace{-0.3mm}(\hspace{-0.2mm}d_{\text{u}_i\text{r}}, \hspace{-0.3mm} \zu\hspace{-0.2mm}) \right] \hspace{-0.7mm}
	\exp \hspace{-0.7mm} \left(\hspace{-0.3mm} - \frac{\gamma_\text{t} \Noise} {\rho_\text{r}} \hspace{-0.3mm} \right) \nonumber \\
	& \quad \times
	\mathbb{E}_{\Phi_\text{e}} 
	\left[\prod_{\text{e}_i \in \Phi_\text{e}} 
	\left\{ 1 - \exp \left( - \frac{\gamma_\text{t}' \Noise} {\rho_{\text{e}_i}} \right) \right.\right. \nonumber \\
	& \quad \left.\left. \times
	\mathbb{E}_{\Phi_\text{u}} \hspace{-1mm}
	\left[	\prod_{\text{u}_i  \in \Phi_\text{u}} 
	\sum_{e_\text{J} \in \{\text{L}, \text{N}\}} \hspace{-1mm}
	\hat{p}_{\text{s,e}_i}^{(e_\text{J})}(d_{\text{u}_i\text{e}_i}, \zu) p_{e_\text{J}}(d_{\text{u}_i\text{e}_i},\zu) \right] \right\} \right] \hspace{-2mm}
	\label{eq:P_network}
	\end{align}
	where (a) is obtained because $h_{\text{tr}} \sim \exp(1)$ and $h_{\text{te}_i} \sim \exp(1)$, and (b) is from the \ac{CDF} of $h_{\text{u}_i{\text{r}}}$ and $h_{\text{u}_i{\text{e}_i}}$.
	By using the \ac{PGFL} in \eqref{eq:P_network}, $\tilde{p}_\text{se}(\zu)$ is presented as \eqref{eq:p_se_multiple}.
\end{IEEEproof}
%
%

In a similar way to the Corollary \ref{pro:asymptotic}, we provide the asymptotic analysis of the secrecy transmission probability for multiple \ac{UAV} jammers.
Specifically, in \eqref{eq:p_se_multiple}, when $\zu$ goes zero, $p_\text{L}(r,\zu)$ and $p_\text{L}(v,\zu)$ approach to zero and the secrecy transmission probability can be represented as
\begin{align}
	\tilde{p}_\text{se} &\approx 
	\exp \left\{
	- 2 \pi \lambda_\text{u} \int_{0}^{\infty} 
	\frac{\gamma_\text{t} P_\text{u} \ell_\text{r}^{\alpha_\text{N}} r }  
	{P_\text{t} r^{\alpha_{\text{N}}} + \gamma_\text{t} P_\text{u} \ell_\text{r}^{\alpha_{\text{N}}}} \, dr 
	- \frac{\gamma_\text{t} \Noise} {P_\text{t} \ell_\text{r}^{-\alpha_{\text{N}}}} \right\}
	\nonumber \\
	& \quad \times
	\exp\left[- 2 \pi \Lw \int_{0}^{\infty} 
	\exp\left\{
	- \frac{\gamma_\text{t}' \Noise \ell_{\text{e}_i}^{\alpha_{\text{N}}}} {P_\text{t}} \right.\right. \nonumber \\
	& \left. \left. \quad
	- 2 \pi \lambda_\text{u} 
	\int_{0}^{\infty}  \frac{\gamma_\text{t}' P_\text{u} \ell_{\text{e}_i}^{\alpha_{\text{N}}} v}  
	{P_\text{t} v^{\alpha_{\text{N}}} + \gamma_\text{t}' P_\text{u} \ell_{\text{e}_i}^{\alpha_{\text{N}}}} \, dv \right\} \hspace{-0.5mm} \ell_{\text{e}_i} d\ell_{\text{e}_i} \right]. \label{eq:p_se_asy2}
\end{align}
where the integral term is represented as \cite[eq. (3.241)]{GraRyz:B07}
\begin{align}
	\int_{0}^{\infty} \hspace{-1mm}\frac{x^{\mu-1}} {\left(p \hspace{-0.7mm} + \hspace{-0.7mm} qx^\nu\right)^{n+1}} dx
	\hspace{-0.5mm} = \hspace{-0.5mm} \frac{1}{\nu p^{n+1}} \hspace{-0.5mm} \left(\frac{p}{q}\right)^{\hspace{-0.5mm}\frac{\mu}{\nu}} \frac{\Gamma\hspace{-0.5mm}\left(\frac{\mu}{\nu}\right) \hspace{-0.5mm} \Gamma\hspace{-0.5mm}\left(1 \hspace{-0.7mm} + \hspace{-0.7mm} n \hspace{-0.7mm} - \hspace{-0.7mm} \frac{\mu}{\nu}\right)} {\Gamma\left(1 + n\right)} \hspace{-1.5mm} \label{eq:integral5}
\end{align}
with $p=\gamma_\text{t} P_\text{u} \ell_\text{r}^{\alpha_\text{N}}$ (or $p=\gamma_\text{t}' P_\text{u} \ell_{\text{e}_i}^{\alpha_\text{N}}$), $q=P_\text{t}$, $\nu=\alpha_\text{N}$, $\mu=2$, and $n=0$.
By using \eqref{eq:integral5} in \eqref{eq:p_se_asy2} and $\Gamma(x)\Gamma(1-x) = \frac{\pi}{\sin(\pi x)}$, $\tilde{p}_\text{se}$ can be expressed as
\begin{align}
	\tilde{p}_\text{se} &\approx \exp \left\{
	- \frac{2 \pi^2 \lambda_\text{u}}{\alpha_\text{N} \sin\left(\frac{2 \pi}{\alpha_\text{N}}\right)} \left(\frac{\gamma_\text{t} \ell_\text{r}^{\alpha_\text{N}} P_\text{u}} {P_\text{t}}\right)^{\frac{2}{\alpha_\text{N}}}
	- \frac{\gamma_\text{t} \Noise} {P_\text{t} \ell_\text{r}^{-\alpha_{\text{N}}}} \right\} \nonumber \\
	& \quad \times
	\exp\left[- 2 \pi \Lw \int_{0}^{\infty} 
	\exp\left\{
	- \frac{\gamma_\text{t}' \Noise \ell_{\text{e}_i}^{\alpha_{\text{N}}}} {P_\text{t}} \right. \right. \nonumber \\ 
	& \quad \left. \left. - \frac{2 \pi^2 \lambda_\text{u}}{\alpha_\text{N} \sin\left(\frac{2 \pi}{\alpha_\text{N}}\right)} \left(\frac{\gamma_\text{t}' P_\text{u}} {P_\text{t}}\right)^{\frac{2}{\alpha_\text{N}}} \ell_{\text{e}_i}^2 \right\} \ell_{\text{e}_i} d\ell_{\text{e}_i} \right] . \label{eq:final}
\end{align}
In \eqref{eq:final}, when $\alpha_\text{N}=4$, by substituting $\ell_{\text{e}_i}^2 = t$,  $\tilde{p}_\text{se}$ is given by
\begin{align}
	\tilde{p}_\text{se} &\approx
	\exp \left\{
	\hspace{-0.7mm} - \frac{\pi^2 \lambda_\text{u}}{2} \left(\frac{\gamma_\text{t} \ell_\text{r}^{4} P_\text{u}} {P_\text{t}}\right)^{\hspace{-0.7mm}\frac{1}{2}}
	\hspace{-0.7mm} - \hspace{-0.7mm} \frac{\gamma_\text{t} \Noise} {P_\text{t} \ell_\text{r}^{-4}} \right\} 
	\exp\left[- \pi \Lw \hspace{-0.7mm} \int_{0}^{\infty}  \right. \nonumber \\
	& \left. \quad \times
	\exp\left\{
	\hspace{-0.7mm} - \frac{\gamma_\text{t}' \Noise t^2} {P_\text{t}} 
	\hspace{-0.7mm} - \hspace{-0.7mm} \frac{\pi^2 \lambda_\text{u}}{2} \left(\frac{\gamma_\text{t}' P_\text{u}} {P_\text{t}}\right)^{\hspace{-0.7mm}\frac{1}{2}} t \right\} \, dt \right] . \label{eq:p_se_asy3}
\end{align}
Using the following result \cite[eq. (3.322)]{GraRyz:B07}
\begin{align}
	\int_{0}^{\infty} \hspace{-1.2mm} \exp\hspace{-0.9mm}\left(\hspace{-0.7mm}- \frac{x^2}{4 \beta} \hspace{-0.5mm} - \hspace{-0.5mm} \gamma x\hspace{-0.8mm}\right) \hspace{-0.8mm} dx
	\hspace{-0.8mm} = \hspace{-0.8mm} \sqrt{\hspace{-0.2mm}\pi \beta} \hspace{-0.3mm} \exp\hspace{-0.8mm}\left(\beta \gamma^2 \right) \hspace{-0.8mm} \left\{\hspace{-0.3mm}1 \hspace{-0.7mm} - \hspace{-0.7mm} \Phi \hspace{-0.8mm}\left(\hspace{-0.3mm}\gamma \sqrt{\beta} \right) \hspace{-0.8mm}\right\} \hspace{-1mm} \label{eq:integral6}
\end{align}
with $\beta=\frac{P_\text{t}}{4 \gamma_\text{t}' \sigma^2}$ and $\gamma=\frac{\pi^2 \lambda_\text{u}}{2} \left(\frac{\gamma_\text{t}' P_\text{u}} {P_\text{t}}\right)^{\frac{1}{2}}$,
$\tilde{p}_\text{se}$ in \eqref{eq:p_se_asy3} can be presented in closed-form as
\begin{align}
\tilde{p}_\text{se} &\approx
\exp \left[
-\frac{\pi^2 \lambda_\text{u} \left(\frac{\gamma_\text{t} \ell_\text{r}^{4} P_\text{u}}{P_\text{t}}\right)^{\frac{1}{2}}} {2} 
-\frac{\gamma_\text{t} \Noise}{P_\text{t} \ell_\text{r}^{-4}}
- \pi \lambda_\text{e} \sqrt{\frac{\pi P_\text{t}}{4 \gamma_\text{t}' \Noise}}  \right. \nonumber \\
& \left. \quad \times
\exp\left(\frac{\pi^4 \lambda_\text{u}^2 P_\text{u}} {16 \Noise}\right)
\left\{1 - \Phi\left(\frac{\pi^2 \lambda_\text{u}}{4} \sqrt{\frac{P_\text{u}}{\Noise}}\right)\right\}
\right]
\end{align}
where $\Phi\left(x\right)=\frac{2}{\sqrt{\pi}}\int_{0}^{\pi} \exp\left(-t^2\right) dt$ is the error function.
From this result, we can see the effect of $\lambda_\text{u}$ on the secrecy transmission probability.

%
%
%
\section{Numerical Results}\label{sec:numerical}

In this section, we evaluate the secrecy transmission probability depending on the location and the transmission power of the Jammer.
Unless otherwise specified, the values of simulation parameters are $\alpha_\text{N}=3.5$, $\alpha_\text{L}=2.5$, $m_\text{L}=2$, $\nu\hspace{-0.5mm}=\hspace{-0.5mm}5\times10^{-4}$, $\mu\hspace{-0.5mm}=\hspace{-0.5mm}0.3$, $\zeta\hspace{-0.5mm}=\hspace{-0.5mm}15$, $R\hspace{-0.5mm}=\hspace{-0.5mm}10000m$, $\gamma_\text{t}=3$, $\gamma_\text{t}'=2.5$, $\Pm=10^{-8}W$, $\PI=3\times10^{-10}W$, and $\Noise=3\times10^{-19}W$.

%
\begin{figure}[t!]
	\begin{center}   
		{ 
			\psfrag{AAAAAAAAAAAAAAAAA1111}[Bl][Bl][0.59]{$\zu=0m$ (Ground Jammer)}
			\psfrag{B1}[Bl][Bl][0.59]{$\zu=100m$}
			\psfrag{C1}[Bl][Bl][0.59]{$\zu=200m$}
			\psfrag{D1}[Bl][Bl][0.59]{Simulation}
			\psfrag{E1}[Bl][Bl][0.59]{$\Lw=5\times10^{-7} [\text{nodes/}m^2]$}
			\psfrag{F1}[Bl][Bl][0.59]{$\Lw=7\times10^{-7} [\text{nodes/}m^2]$}
			\psfrag{G1}[Bl][Bl][0.59]{Asymptotic analysis}
			\psfrag{H1}[Bl][Bl][0.59]{$m_\text{L}=2$}
			\psfrag{J1}[Bl][Bl][0.59]{$m_\text{L}=9$}
			\psfrag{X1}[Bl][Bl][0.75]{Horizontal distance between Tx and Jammer, $\dk \:[m]$}
			\psfrag{Y1}[Bl][Bl][0.75]{Secrecy Transmission Probability, $p_\text{se}(\dk,\zu,\pi)$}
			\includegraphics[width=0.9\columnwidth]{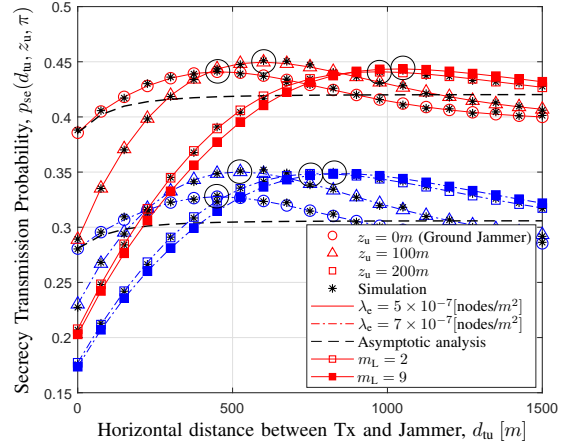}
			\vspace{-4.5mm}
		}
	\end{center}
	\caption{
		Secrecy transmission probabilities $p_\text{se}(\dk,\zu,\pi)$ as a function of $\dk$ with $\om=340m$ for different values of $\Lw$ and $\zu$.
		The optimal values of $\dk$ are marked by circles.
		\vspace{-3.5mm}
	}
	\label{fig:secrecy1}
\end{figure}
%

Figure \ref{fig:secrecy1} presents the secrecy transmission probability $p_\text{se}(\dk,\zu,\pi)$ as a function of the horizontal distance between the Jammer and the Tx $\dk$ with $\om=340m$ for different values of the Eve density $\Lw$ and the Jammer height $\zu$.
From Fig. \ref{fig:secrecy1},
we can see that $p_\text{se}(\dk,\zu,\pi)$ first increases with $\dk$ up to a certain value of $\dk$, and then decreases.
This is because
for small $\dk$, the decrease in the \ac{LoS} probability of the interference link to the receiver is greater than that of the jamming link to the Eve with $\max\limits_{\text{e}_i \in \Phi_{\text{e}}}  \gamma_{\text{e}_i}$, who mainly affects the eavesdropping probability.
On the other hand, for large $\dk$, as $\dk$ increases, the Eve with $\max\limits_{\text{e}_i \in \Phi_{\text{e}}}  \gamma_{\text{e}_i}$ can be located closer to the Tx than the Rx, so $p_\text{se}(\dk,\zu,\pi)$ decreases with $\dk$.
%
We can also see that as $\Lw$ increases, the optimal value of $\dk$ decreases to make the jamming link stronger as there exist more Eves.
From this, we can find out that as the density of Eves increases, the Jammer needs to be located nearer to the Tx.
In Fig.~\ref{fig:secrecy1}, we can additionally see the impact of $m_\text{L}$ on $\Pse$ according to $d_\text{tu}$.
	Specifically, for small $d_\text{tu}$, $\Pse$ decreases with $m_\text{L}$ 
	since $\Pstp$ decreases with $m_\text{L}$ more than $\Peve$.
	On the other hand, for large $d_\text{tu}$, $\Pse$ increases with $m_\text{L}$ 
	since $\Pstp$ becomes similar for different $m_\text{L}$, but $\Peve$ still decreases with $m_\text{L}$.
	Hence, the optimal value of $d_\text{tu}$ increases with $m_\text{L}$, 
	which means the Jammer needs to be located further from the Tx as $m_\text{L}$ increases.
Furthermore, we can know that the asymptotic analysis almost matches the analytic analysis as the Jammer approaches to the Tx (i.e., as $\dk \rightarrow 0$ for $\zu=0$).

\begin{figure}[t!]
	\begin{center}   
		{ 
			\psfrag{AAAAAAAAAAAAAAAAAAA11}[Bl][Bl][0.59]{$\Lw=1.2\times10^{-8} [\text{nodes/}m^2]$}
			\psfrag{B}[Bl][Bl][0.59]{$\Lw=7\times10^{-8} [\text{nodes/}m^2]$}
			\psfrag{C}[Bl][Bl][0.59]{$\Lw=3.5\times10^{-7} [\text{nodes/}m^2]$}
			\psfrag{D}[Bl][Bl][0.59]{$\Lw=7.5\times10^{-7} [\text{nodes/}m^2]$}
			\psfrag{E}[Bl][Bl][0.59]{$\Lw=9.3\times10^{-7} [\text{nodes/}m^2]$}
			\psfrag{X}[Bl][Bl][0.75]{Height of Jammer, $\zu \:[m]$}
			\psfrag{Y}[Bl][Bl][0.75]{Horizontal distance between Tx and Jammer, $\dk \:[m]$}
			\includegraphics[width=0.9\columnwidth]{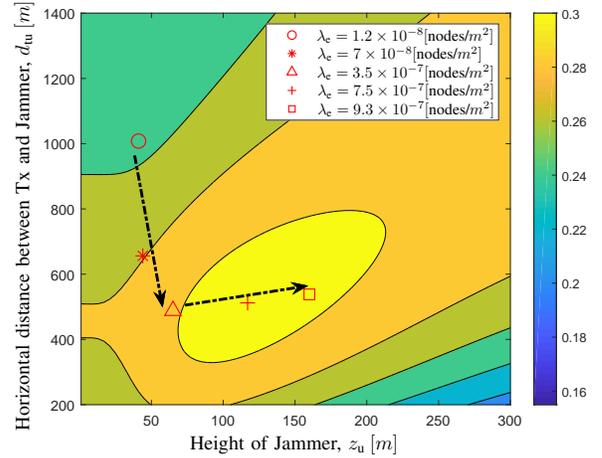}
			\vspace{-4.5mm}
		}
	\end{center}
	\caption{
		Secrecy transmission probability $p_\text{se}(\dk,\zu,\pi)$ as functions of $\zu$ and $\dk$ with $\Lw\hspace{-0.7mm}=\hspace{-0.7mm}7.5\hspace{-0.7mm}\times\hspace{-0.7mm}10^{-7} [\text{nodes/}m^2]$ and $\om\hspace{-0.7mm}=\hspace{-0.7mm}420m$. The optimal Jammer locations for each $\Lw$ that maximize $p_\text{se}(\dk,\zu,\pi)$ are marked by symbols.
		\vspace{-5mm}
	}
	\label{fig:secrecy3}
\end{figure}
%

Figure \ref{fig:secrecy3} presents the secrecy transmission probability $p_\text{se}(\dk,\zu,\pi)$ as functions of the Jammer height $\zu$ and the horizontal distance between the Tx and the Jammer $\dk$ with $\om=420m$.
The symbols mean the optimal Jammer locations, $\zu^*$ and $\dk^*$, for each Eve density $\Lw$.
From Fig. \ref{fig:secrecy3},
we can see that $\dk^*$ first decrease with $\Lw$ up to a certain value of $\Lw$, and then increase.
This is because, for small $\Lw$ (e.g., $\Lw \le 7\times10^{-8} \text{nodes/}m^2$), since there exist less eavesdroppers, the Jammer needs to be located at the low height to reduce the \ac{LoS} probability of interference link to the Rx.
However, for relatively high $\Lw$ (e.g., $\Lw=3.5\times10^{-7} \text{nodes/}m^2$), the Jammer needs to be located closer to the Tx, especially by reducing the horizontal distance $\dk$ for giving stronger jamming signal to Eves, although it also gives larger interference to the Rx.
Additionally, when $\Lw$ is much higher (e.g., $\Lw \ge 7.5\times10^{-7} \text{nodes/}m^2$), since there exist many Eves, the Jammer needs to give much stronger jamming signal to Eves.
Hence, the Jammer is located at the high height to increase the \ac{LoS} probability of the jamming link.

%
\begin{figure}[t!]
	\begin{center}   
		{ 
			\psfrag{AAAAAAAAAAAAAAAAAA11}[Bl][Bl][0.59]{$\lambda_\text{u}=7\times10^{-6} [\text{nodes/}m^2]$}
			\psfrag{B1}[Bl][Bl][0.59]{$\lambda_\text{u}=9\times10^{-6} [\text{nodes/}m^2]$}
			\psfrag{C1}[Bl][Bl][0.59]{$\PI=2\times10^{-11}W$}
			\psfrag{D1}[Bl][Bl][0.59]{$\PI=3\times10^{-11}W$}
			\psfrag{X1}[Bl][Bl][0.75]{Height of UAV jammers, $\zu \:[m]$}
			\psfrag{Y1}[Bl][Bl][0.75]{$\tilde{p}_\text{se}(\zu)$}
			\includegraphics[width=0.9\columnwidth]{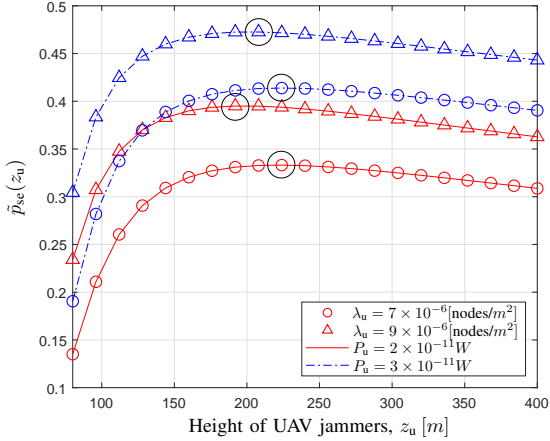}
			\vspace{-4.5mm}
		}
	\end{center}
	\caption{
		Secrecy transmission probability $\tilde{p}_\text{se}(\zu)$ as a function of $\zu$ for different values of $\lambda_\text{u}$ and $\PI$. The optimal values of $\zu$ are marked by circles.
		\vspace{-10mm}
	}
	\label{fig:secrecy4}
\end{figure}
%

Figure \ref{fig:secrecy4} presents the secrecy transmission probability $\tilde{p}_\text{se}(\zu)$ as a function of the height of \ac{UAV} jammers $\zu$ for different values of the \ac{UAV} jammer density $\lambda_\text{u}$ and the transmission power of the \ac{UAV} jammer.
Here, we use $\om=50m$ and $\Lw=10^{-5} \text{nodes/}m^2$.
From Fig. \ref{fig:secrecy4}, 
we can see that $\tilde{p}_\text{se}(\zu)$ first increases with $\zu$ up to a certain value of $\zu$, and then decreases.
This is because
for small $\zu$, the increase in the \ac{LoS} probability of the jamming link is greater than that of the interference link.
On the other hand, for large $\zu$, the \ac{LoS} probability of the interference link keeps increasing, while the distance-dependent path loss of the eavesdropping link decreases.
Therefore, $\tilde{p}_\text{se}(\zu)$ decreases with $\zu$ when large $\zu$.
We can also see that as $\lambda_\text{u}$ increases, the optimal value of $\zu$ decreases to give weaker \ac{LoS} probability (i.e., weaker signal) on the interference link to the Rx.
Furthermore, the optimal value of $\zu$ increases as $\PI$ increases.
From these results, we can know that when the effect of \ac{UAV} jammers is strong enough by using the larger transmit power, the \ac{UAV} jammers need to be located at the low height to reduce the \ac{LoS} probability of the interfering signal at the Rx or located at the high height to decrease the distance-dependent path loss of the interference link.

%
%
\section{Conclusion}\label{sec:conclusion}
This paper derives and analyzes the secrecy transmission probability of \ac{UAV} communications considering the realistic channel models affected by the communication link.
Using the derived expression, we show the effect of a \ac{UAV} friendly jammer on network parameters.
Specifically, as the \ac{UAV} height increases, the distance-dependent path loss decreases, but the \ac{LoS} probability for jamming signal increases.
From this relation, we show that there can exists an optimal \ac{UAV} height, which decreases as the density of \ac{UAV} jammers increases for the multiple Jammer case.
We also provide that as the Eve density increases or the Jammer height becomes lower, the optimal horizontal distance between the Jammer and the transmitter decreases to make the jamming link stronger.
The outcomes of this work can provide insights on the optimal deployment of the friendly \ac{UAV} jammer that prevents eavesdropping while reducing the interference to the receiver.
%
%
%
%
%

\bibliographystyle{IEEEtran}

\bibliography{StringDefinitions,IEEEabrv,mybib}

\end{document}